\documentclass{pasa}%

\title[A Plane-Parallel Wind Solution]{A Plane-Parallel Wind Solution For Testing Numerical Simulations of Photoevaporation}
\author[Hutchison \& Laibe]{
Mark A. Hutchison$^{1}$\thanks{E-mail: mhutchison@swin.edu.au}
and Guillaume Laibe$^{2}$
\\
\affil{$^{1}$Centre for Astrophysics \& Supercomputing, Swinburne University of Technology, Hawthorn, VIC 3122, Australia}%
\affil{$^{2}$School of Physics and Astronomy, University of St. Andrews, North Haugh, St. Andrews, Fife KY16 9SS, UK}}%
\jid{PASA}
\doi{10.1017/pas.\the\year.xxx}
\jyear{\the\year}

\usepackage[authoryear]{natbib}
\bibpunct{(}{)}{;}{a}{}{,}
\usepackage{aas_macros}
\usepackage{hyperref} 
\hypersetup{colorlinks,citecolor=blue,linkcolor=blue,urlcolor=blue}

\usepackage[capitalise]{cleveref}
     \crefname{equation}{equation}{equations}
     \crefname{figure}{figure}{figures}
     \crefname{table}{table}{tables}
     
\usepackage{enumerate}
\usepackage{enumitem}
\usepackage{flushend}
\usepackage{tabto}

\usepackage[usenames, dvipsnames]{color}

\begin{document}%
\begin{abstract}
Here we derive a Parker-wind like solution for a stratified, plane-parallel atmosphere undergoing photoionisation. The difference compared to the standard Parker solar wind is that the sonic point is crossed only at infinity. The simplicity of the analytic solution makes it a convenient test problem for numerical simulations of photoevaporation in protoplanetary discs.
\end{abstract}
\maketitle%

\section{INTRODUCTION }
\label{sec:intro}

Photoevaporation is a pressure-driven wind produced by high energy stellar radiation that heats and/or ionises gas located in the incident surface layers of protoplanetary atmospheres \citep{Hollenbach/etal/1994}. If the thermal energy of the heated gas exceeds the gravitational binding energy of the central gravitating body, the gas is unbound and can escape in a slow, often centrifugally launched, wind. These winds are similar in nature to the familiar pressure-driven Parker winds in stars \citep{Parker/1958}, but are made complicated by rotation, disc geometry, and/or off-axis radiation sources. For example, the flow solution for photoionised disc winds cannot rigorously be solved analytically because the solution depends on knowing \emph{a priori} the exact streamline trajectories \citep[or divergence; see][]{Begelman/McKee/Shields/1983}. While trivial for spherically symmetric winds, the extension to discs can only be approximated \citep[e.g.][]{Waters/Proga/2012}.

The analytic solution for isothermal Parker winds has typically been used as a numerical test for hydrodynamic simulations involving astrophysical winds \citep[e.g.][]{Keppens/Goedbloed/1999,Font/etal/2004}. However, apart from sharing a similar transonic wind structure, stellar winds and photoevaporation in discs are physically quite different (e.g. geometry, gravity, temperature, density). If one is only interested in photoevaporating discs, the numerical overhead of setting up alternate conditions necessary to produce stellar winds can be inconvenient. In such cases, it would be ideal to have an analytic solution to a problem that uses the same numerical setup and physical parameters as a real disc.

An analytic wind solution for photoevaporation in a \emph{disc}-like environment can be derived using a non-rotating, stratified, plane-parallel atmosphere. On local scales, the vertical structure of protoplanetary discs is approximately plane-parallel so the physical parameters and numerical setup can be made to be almost identical to that of a disc at any given radius. The resulting wind's simple 1-D geometry makes the solution analytically tractable and straight forward to use as an alternative test to the isothermal Parker wind---its utility has motivated this study.

\section{ANALYTIC FLOW SOLUTION}
\label{sec:analytic_solution}

The relevant equations describing a steady-state, pressure-driven, isothermal Parker wind come from setting $\partial/\partial t = 0$ in the fluid equations:
\begin{align}
	\nabla \cdot (\rho \mathbf{v}) & = 0,
	\label{eq:continuity}
\\
	\rho \left( \mathbf{v} \cdot \nabla \mathbf{v} \right) & =  - \nabla P + \rho \mathbf{g},
	\label{eq:momentum}
\\
	P & = \rho \mathcal{R} T,
	\label{eq:eos}
\\
	T & = T_0,
	\label{eq:isothermal}
\end{align}
where $\mathbf{g}$ is the gravitational force, $\mathcal{R}$ is the gas constant, and $\rho$, $\mathbf{v}$, $P$, and $T$ are the gas density, velocity, pressure, and temperature, respectively. In anticipation of applying this test to photoevaporating circumstellar discs, we define the gravity $\mathbf{g}$ to be the \emph{vertical} field produced by a massive central object,
\begin{equation}
	\mathbf{g} = -\frac{\mathcal{G} M z}{(R^2+z^2)^{3/2}}\mathbf{\hat{z}},
	\label{eq:vertical_gravity}
\end{equation}
to ensure a disc-like density and temperature structure in the atmosphere. Here $\mathcal{G}$ is the gravitational constant, $M$ is the mass of the central star, and $R$ is the cylindrical distance from the central source to our local patch of atmosphere. Without loss of generality, we restrict the variables to be functions of $z$ only. To close the set of equations, we adopt the equation of state of an ideal gas ($pV = n\mathcal{R}T=$ constant, where $n$ is the number of moles of the gas) such that the sound speed of the wind is constant and can be written as $c_\text{s}^2 = P/\rho$.

Integrating \cref{eq:continuity} gives $\rho v = \text{constant}$, or written in terms of an accretion rate \citep{Bondi/1952}, 
\begin{equation}
	\dot{M} = A \rho v,
	\label{eq:accretion_rate}
\end{equation}
where $A$ is a problem dependent characteristic surface area.
Meanwhile, using the sound speed relationship to replace $P$, \cref{eq:momentum} can be rewritten as,
\begin{equation}
	v \frac{\text{d} v}{\text{d} z} = - \frac{c_\text{s}^2 }{\rho} \frac{\text{d}\rho}{\text{d} z} - \frac{\mathcal{G} M z}{(R^2+z^2)^{3/2}}.
	\label{eq:1d_momentum}
\end{equation}
The dependence here on $\rho$ can be removed by taking the derivative of \cref{eq:accretion_rate}. After some manipulation we obtain,
\begin{equation}
	-\frac{1}{\rho} \frac{\text{d}\rho}{\text{d}z} = \frac{1}{v} \frac{\text{d}v}{\text{d}z},
\end{equation}
which can immediately be substituted back into \cref{eq:1d_momentum} to obtain,
\begin{equation}
	v \frac{\text{d} v}{\text{d} z}  = \frac{c_\text{s}^2 }{v} \frac{\text{d} v}{\text{d} z} - \frac{\mathcal{G} M z}{(R^2+z^2)^{3/2}}.
\end{equation}
Collecting the derivatives on $v$ and using the following relation,
\begin{equation}
	v \frac{\text{d} v }{\text{d}z} = \frac{ c_\text{s}^2}{2} \frac{\text{d} (v^2/c_\text{s}^2)}{\text{d}z},
\end{equation}
we obtain a separable, ordinary differential equation for $v^2/c_\text{s}^2$:
\begin{equation}
	\left( 1-\frac{c_\text{s}^2}{v^2}  \right)  \frac{\text{d} (v^2/c_\text{s}^2)}{\text{d}z} = - \frac{2 \mathcal{G} M z}{c_\text{s}^2 \, (R^2+z^2)^{3/2}}.
	\label{eq:ode_with_dim}
\end{equation}
Nondimensionalising \cref{eq:ode_with_dim} using $\bar{v}^2 \equiv v^2/c_\text{s}^2$, $\bar{z} \equiv z/R$, the Keplerian Mach number $\mathcal{M} \equiv v_\text{K}/c_\text{s}$, and $v_\text{K} = \sqrt{\mathcal{G} M/R}$, we obtain,
\begin{equation}
	\left( 1-\frac{1}{\bar{v}^2}  \right)  \frac{\text{d} \left(\bar{v}^2\right)}{\text{d}\bar{z}} = - \frac{2 \mathcal{M}^2 \bar{z}}{\left(1+\bar{z}^2\right)^{3/2}}.
	\label{eq:ode}
\end{equation}
Note the presence of a critical point located at the sonic point, $\bar{v} = 1$, on the left-hand side of the equation. From inspection of the right-hand side, the corresponding position must be at $|\bar{z}| \rightarrow \infty$. For comparison, the spherically symmetric isothermal Parker-wind solution is \emph{transonic} with the sonic point located at $r_\text{s} \equiv \mathcal{G}M/2c_\text{s}^2$.

\begin{figure*}
	\centering{\includegraphics[width=0.89\textwidth]{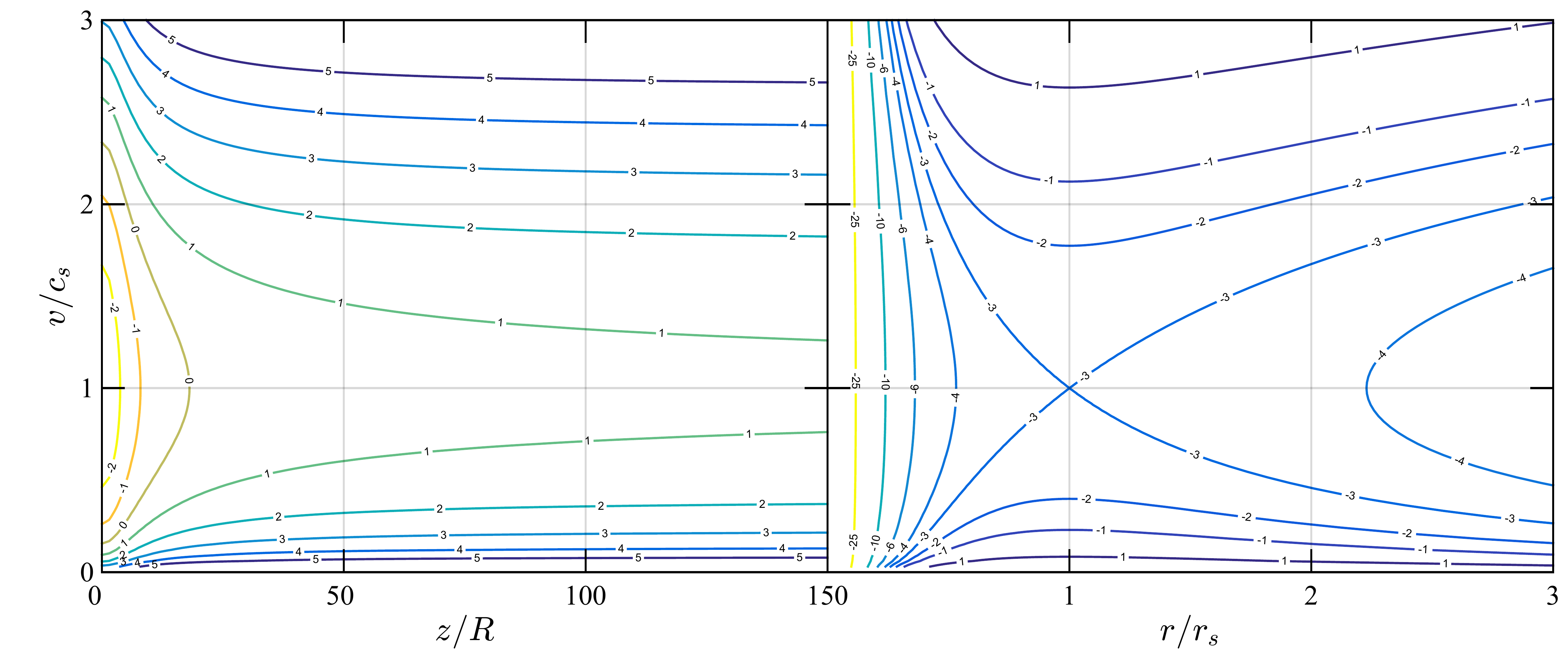}}
	\caption{Comparison between the plane-parallel wind at $R=5\,$AU (left) derived in this paper and the more familiar spherically-symmetric Parker wind (right). The different contour levels are determined by the value of $C$ in \cref{eq:plane-parallel_solution}. The sonic point for the plane-parallel case is only asymptotically crossed as $z \rightarrow \infty$ whereas the Parker Wind model is transonic at $r_\text{s} = \mathcal{G}M/2c_\text{s}^2$.}
	\label{fig:compare_winds}
\end{figure*}

Integrating \cref{eq:ode} we obtain a transcendental equation for the outflow velocity as a function of $\bar{z}$,
\begin{equation}
	\bar{v}^2 - \ln{ \bar{v}^2 } = \frac{2\mathcal{M}^2}{\sqrt{1+\bar{z}^2}} + C,
\end{equation}
where $C$ is an integration constant. Following \citet{Cranmer/2004}, we can write the solution for the velocity in closed form using the Lambert~$\mathrm{W}$ function \citep{Corless/etal/1996,Veberic/2012}:
\begin{equation}
	\bar{v}^2 = - \mathrm{W}_k \! \! \left[ -\exp{ \left( -\frac{2\mathcal{M}^2}{\sqrt{1+\bar{z}^2}} -C \right)  }   \right],
	\label{eq:plane-parallel_solution}
\end{equation}
where
\begin{equation}
	k = \begin{cases}
		0, & \quad \text{if } \bar{v} \leq 1		
		\\
		-1, & \quad \text{if } \bar{v} > 1.		
	\end{cases}
	\label{eq:lambert_k_value}
\end{equation}
For comparison, the velocity for the spherically-symmetric Parker wind is,
\begin{equation}
	\bar{v}^2 = - \mathrm{W}_k \! \! \left[ - \frac{1}{\bar{r}^{4}}  \exp{ \left( -\frac{4}{\bar{r}} -C \right)  }   \right],
\end{equation}
where $\bar{r} \equiv r/r_\text{s}$ and $r$ is the spherical radius measured from the centre of mass $M$. \Cref{fig:compare_winds} contrasts the two solutions above. As the plane-parallel wind cannot support a finite sonic point without diverging streamlines \citep{Begelman/McKee/Shields/1983}, it looks similar to an isothermal Parker wind with its sonic point remapped to infinity. Consequently, the plane-parallel ``breeze'' solutions (always subsonic) are not hydrostatic at infinity. Another minor difference is that the plane-parallel solutions remain finite at $z=0$ due to having a finite gravitational potential at the midplane of the disc. As a final point of interest, the rate of convergence of $v \rightarrow c_\text{s}$ in the asymptotically transonic solution ($C=1$) can more conveniently be expressed by expanding \cref{eq:plane-parallel_solution} in a Taylor series in the limit $|\bar{z}| \rightarrow \infty$,
\begin{equation}
	\bar{v} \approx 1-\frac{\mathcal{M}}{\sqrt{|\bar{z}|}} + \mathcal{O}\left( \frac{1}{z}\right),
\end{equation}
which, due to the $\sqrt{\bar{z}}$ dependence, makes convergence very slow.

The plane-parallel wind has only three possible classes of solutions:
\begin{enumerate}[label=(\roman*),leftmargin=57pt,labelindent=0pt,itemindent=-32.5pt]
	\item $C<1$: \tabto{0mm} $v(z)$ is double-valued on ${z_i \leq z\leq z_\text{max}}$.
	\item $C=1$: \tabto{0mm} $v(z)$ is \emph{asymptotically} transonic and monotonically increasing for outflow ($k=0$) or decreasing for inflow ($k=-1$).
	\item $C>1$: \tabto{0mm} $v(z)$ is not transonic and monotonically increasing/decreasing.
\end{enumerate}
Physically, the solution should be locally mono-valued for stability while continuity and symmetry of the disc indicate that the velocity should be stationary at $z=0$. Meanwhile, studies of the Parker wind show its breeze solutions to be unstable \citep{Velli/1994}, a result which holds in the zero-curvature limit. The only admissible solution remaining is $C=1$ with $k=0$, i.e. 
\begin{equation}
	\bar{v} = \sqrt{- \mathrm{W}_0 \! \! \left[ -\exp{ \left( -\frac{2\mathcal{M}^2}{\sqrt{1+\bar{z}^2}} -1 \right)  }   \right]},
	\label{eq:final_solution}
\end{equation}
but this too is only marginally stable to Velli's global stability criterion \citep{Velli/1994,Grappin/Cavillier/Velli/1997,Del-Zanna/Velli/Londrillo/1998}. We must therefore turn to numerical simulations to verify the stability of the solution, as suggested by \citet{Waters/Proga/2012}.

\section{NUMERICAL STABILITY}
\label{sec:numerical_stability}

Using \cref{eq:accretion_rate,eq:final_solution} and the equation of state, all of the fluid parameters are uniquely determined. A practical setup for our proposed test can be achieved in three steps:

(i) In a 2- or 3-D box with periodic horizontal boundary conditions, set up a vertically-isothermal atmosphere using the thin-disc approximation and the gravitational force given in \cref{eq:vertical_gravity}.

(ii) Instantaneously heat any fluid that falls below some density threshold to a high temperature \citep[e.g. $T=10\,000\,$K to mimic ultraviolet photoevaporation; see][]{Alexander/Clarke/Pringle/2006a}.

(iii) Create a steady-state flow using a vertical boundary condition appropriate for the numerical method of choice. In smoothed particle hydrodynamics (SPH), this is done with a \emph{dynamic} vertical boundary condition that is constrained to move at the prescribed analytic velocity from \cref{eq:final_solution}. The benefit of this method is that steady-state solution is obtained almost immediately. Grid based codes, on the other hand, will typically converge to the steady-state solution using fixed outflow boundary conditions. However, if convergence is too slow, dynamically forcing a small section of the outflow until it exits the computational domain will help precipitate steady-state flow.

\begin{figure*}
	\centering{\includegraphics[width=0.89\textwidth]{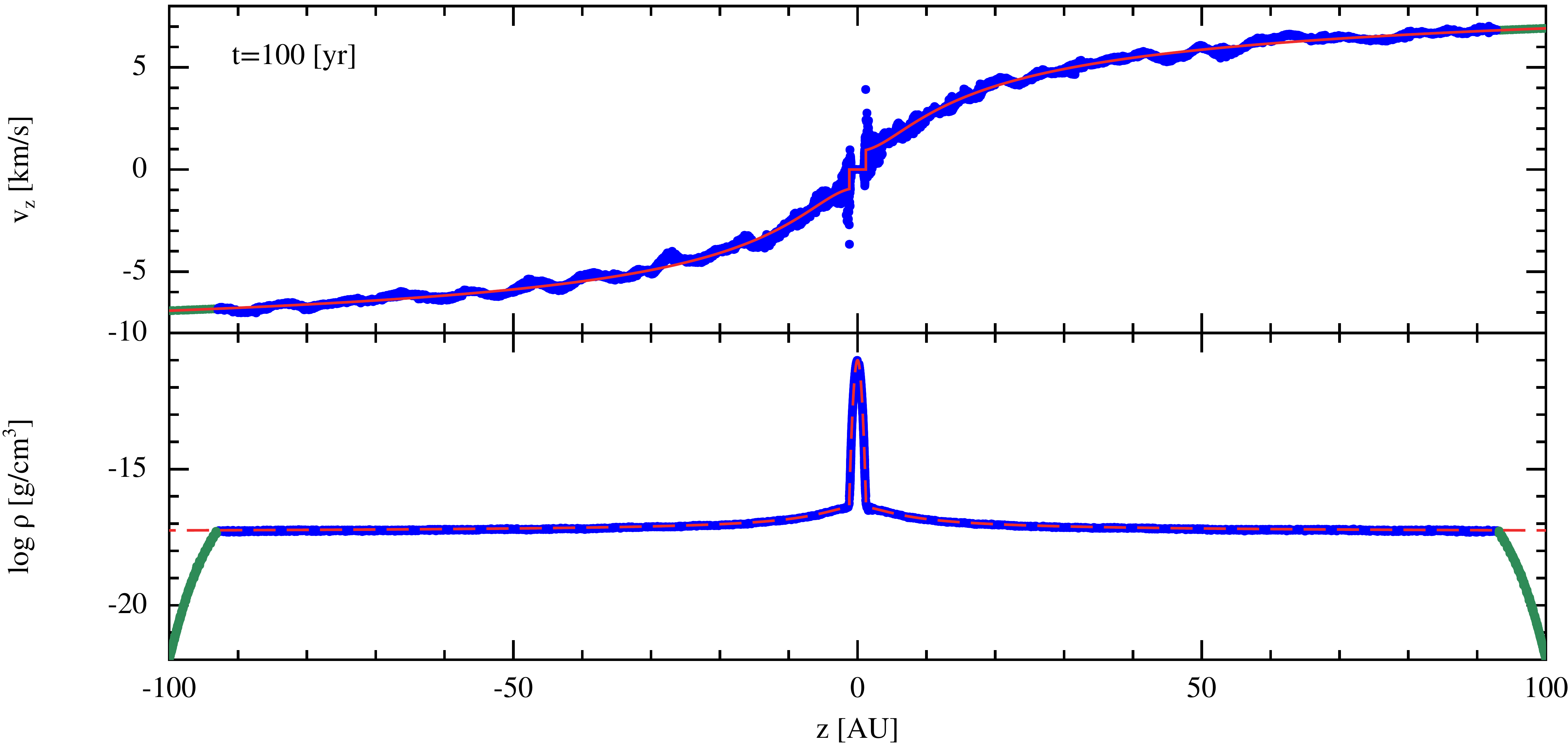}}
	\caption{Velocity (top) and density (bottom) for $200\,028$ SPH gas particles plotted against their respective analytic solutions (red solid and dashed lines) in a two-dimensional plane-parallel disc wind. The green points make up a semi-uniform lattice of gas particles that form a moving boundary condition constrained to move at the velocity prescribed by the analytic solution in \cref{eq:final_solution}. The blue points are the unrestrained gas particles. The $L_2$ errors between the analytic and the numerical solutions are $< 2\%$, consistently with the second-order SPH scheme used.}
	\label{fig:analytic_photo_test}
\end{figure*}

Implementing the setup and SPH boundary conditions described above, we perform the photoevaporation test using our SPH code \textsc{gdphoto} \citep{Hutchison/etal/2016b}. \textsc{Gdphoto} has been benchmarked against the test suite described in \citet{Laibe/Price/2012a} and achieves accuracies comparable to commonly used SPH codes. Using $200\,028$ particles we create a 2-D disc in isothermal hydrostatic equilibrium with the following physical parameters: $M = 1\,M_\odot$, $R=5\,$AU, and $\rho_0=10^{-11}\,$g/cm$^3$. We then initiate photoevaporation by ionising all particles with densities that are five orders of magnitude below the disc midplane density. Ionised particles are held isothermally at $T=10\,000\,$K such that $c_\text{s} \approx 10\,$km/s. \Cref{fig:analytic_photo_test} shows the results after $100\,$yr plotted together with the analytic solution from \cref{eq:final_solution}. The $L_2$ errors for the velocity and density, computed using \textsc{splash} \citep{Price/2007}, are $\sim2$ and $\lesssim 1 \%$, respectively.

\section{DISCUSSION}
\label{sec:conclusion}

The plane-parallel flow described in this paper is comparable to the flow derived by \citet{Adams/2011} for magnetically controlled outflows from hot Jupiters when the stellar magnetic field completely dominates over the magnetic field produced by the planet (i.e. their $\beta \rightarrow \infty$ limit). The main difference between our analytic solutions stems from assuming different gravitational potentials; however, to apply their solution to photoevaporating discs would require readers to rederive the equations themselves. The solution in this paper is significantly more transparent and its closed form makes it especially easy to implement as a numerical test.

Although we focus on using our plane-parallel model as a numerical test, it may have use in wider applications as well. For example, the model's simple geometry, accurate approximation of winds close to the disc, and closed form wind solution could make it a perfect springboard for developing an analytic or semi-analytic model for coupled two-phase photoevaporation. To date, few studies have focused on dust dynamics in winds and a simple two-phase model would be very valuable.

\section*{Acknowledgements} 

We thank Daniel Price and Sarah Maddison for useful discussions and the anonymous referee whose careful report significantly improved this paper and pointed out the explicit form of the solution.  M.H.\ acknowledges funding from a Swinburne University Postgraduate Research Award (SUPRA). G.\ Laibe acknowledges funding from the European Research Council for the FP7 ERC advanced grant project ECOGAL. The numerical calculations were performed on the SwinSTAR supercomputer at Swinburne University of Technology and the subsequent visualisation was made using \textsc{splash} \citep{Price/2007}.


\bibliographystyle{apj}
\bibliography{$HOME/Dropbox/Bibtex_library/library}

\begin{thebibliography}{18}
\expandafter\ifx\csname natexlab\endcsname\relax\def\natexlab#1{#1}\fi

\bibitem[{{Adams}(2011)}]{Adams/2011}
{Adams}, F.~C. 2011, \apj, 730, 27

\bibitem[{{Alexander} {et~al.}(2006){Alexander}, {Clarke}, \&
  {Pringle}}]{Alexander/Clarke/Pringle/2006a}
{Alexander}, R.~D., {Clarke}, C.~J., \& {Pringle}, J.~E. 2006, \mnras, 369, 216

\bibitem[{{Begelman} {et~al.}(1983){Begelman}, {McKee}, \&
  {Shields}}]{Begelman/McKee/Shields/1983}
{Begelman}, M.~C., {McKee}, C.~F., \& {Shields}, G.~A. 1983, \apj, 271, 70

\bibitem[{{Bondi}(1952)}]{Bondi/1952}
{Bondi}, H. 1952, \mnras, 112, 195

\bibitem[{Corless {et~al.}(1996)Corless, Gonnet, Hare, Jeffrey, \&
  Knuth}]{Corless/etal/1996}
Corless, R., Gonnet, G., Hare, D., Jeffrey, D., \& Knuth, D. 1996, Advances in
  Computational Mathematics, 5, 329

\bibitem[{{Cranmer}(2004)}]{Cranmer/2004}
{Cranmer}, S.~R. 2004, American Journal of Physics, 72, 1397

\bibitem[{{Del Zanna} {et~al.}(1998){Del Zanna}, {Velli}, \&
  {Londrillo}}]{Del-Zanna/Velli/Londrillo/1998}
{Del Zanna}, L., {Velli}, M., \& {Londrillo}, P. 1998, \aap, 330, L13

\bibitem[{{Font} {et~al.}(2004){Font}, {McCarthy}, {Johnstone}, \&
  {Ballantyne}}]{Font/etal/2004}
{Font}, A.~S., {McCarthy}, I.~G., {Johnstone}, D., \& {Ballantyne}, D.~R. 2004,
  \apj, 607, 890

\bibitem[{{Grappin} {et~al.}(1997){Grappin}, {Cavillier}, \&
  {Velli}}]{Grappin/Cavillier/Velli/1997}
{Grappin}, R., {Cavillier}, E., \& {Velli}, M. 1997, \aap, 322, 659

\bibitem[{{Hollenbach} {et~al.}(1994){Hollenbach}, {Johnstone}, {Lizano}, \&
  {Shu}}]{Hollenbach/etal/1994}
{Hollenbach}, D., {Johnstone}, D., {Lizano}, S., \& {Shu}, F. 1994, \apj, 428,
  654

\bibitem[{{Hutchison} {et~al.}(2016){Hutchison}, {Price}, {Laibe}, \&
  {Maddison}}]{Hutchison/etal/2016b}
{Hutchison}, M.~A., {Price}, D.~J., {Laibe}, G., \& {Maddison}, S.~T. 2016,
  submitted to MNRAS

\bibitem[{{Keppens} \& {Goedbloed}(1999)}]{Keppens/Goedbloed/1999}
{Keppens}, R. \& {Goedbloed}, J.~P. 1999, \aap, 343, 251

\bibitem[{{Laibe} \& {Price}(2012)}]{Laibe/Price/2012a}
{Laibe}, G. \& {Price}, D.~J. 2012, \mnras, 420, 2345

\bibitem[{{Parker}(1958)}]{Parker/1958}
{Parker}, E.~N. 1958, \apj, 128, 664

\bibitem[{{Price}(2007)}]{Price/2007}
{Price}, D.~J. 2007, \pasa, 24, 159

\bibitem[{{Veberi{\v c}}(2012)}]{Veberic/2012}
{Veberi{\v c}}, D. 2012, Computer Physics Communications, 183, 2622

\bibitem[{{Velli}(1994)}]{Velli/1994}
{Velli}, M. 1994, \apjl, 432, L55

\bibitem[{{Waters} \& {Proga}(2012)}]{Waters/Proga/2012}
{Waters}, T.~R. \& {Proga}, D. 2012, \mnras, 426, 2239

\end{thebibliography}

\end{document}